# Exploring Advanced Methodologies in Security Evaluation for Large Language Models*

Jun Huang[1], Jiawei Zhang[1], Qi Wang[2], Weihong Han[1], and Yanchun Zhang[1]

[1] Peng Cheng Laboratory, Shenzhen, China
{huangj02,zhangjw01,hanwh}@pcl.ac.cn,yanchun.zhang@vu.edu.au
[2] Information Science and Technology College, Dalian Maritime university, Dalian, China
qiwang1101@163.com

**Abstract.** Large Language Models (LLMs) represent an advanced evolution of earlier, simpler language models. They boast enhanced abilities to handle complex language patterns and generate coherent text, images, audios, and videos. Furthermore, they can be fine-tuned for specific tasks. This versatility has led to the proliferation and extensive use of numerous commercialized large models. However, the rapid expansion of LLMs has raised security and ethical concerns within the academic community. This emphasizes the need for ongoing research into security evaluation during their development and deployment. Over the past few years, a substantial body of research has been dedicated to the security evaluation of large-scale models. This article an in-depth review of the most recent advancements in this field, providing a comprehensive analysis of commonly used evaluation metrics, advanced evaluation frameworks, and the routine evaluation processes for LLMs. Furthermore, we also discuss the future directions for advancing the security evaluation of LLMs.

## 1 Introduction

Language models (LMs) are statistical machines designed to comprehend natural language, which calculate the likelihood of word sequences based on vast collections of text [5,58].

Large language models (LLMs) represent a significant advancement over their simpler predecessors, showcasing enhanced abilities to understand intricate linguistic patterns and generate coherent, contextually appropriate text [1,45,55]. In addition to text, LLMs can also process pictures [3], audio recordings [4], and videos [2]. Moreover, they can be fine-tuned to excel at specific tasks with utmost precision. With the impressive growth of LLMs' capabilities, numerous commercially available large models, such as OpenAI's GPT series, Anthropic's Claude series, Google's T5 and PaLM, Meta's OPT, and Baidu's Ernie [57], have flourished and found extensive applications.

---

* Supported by Peng Cheng Laboratory.

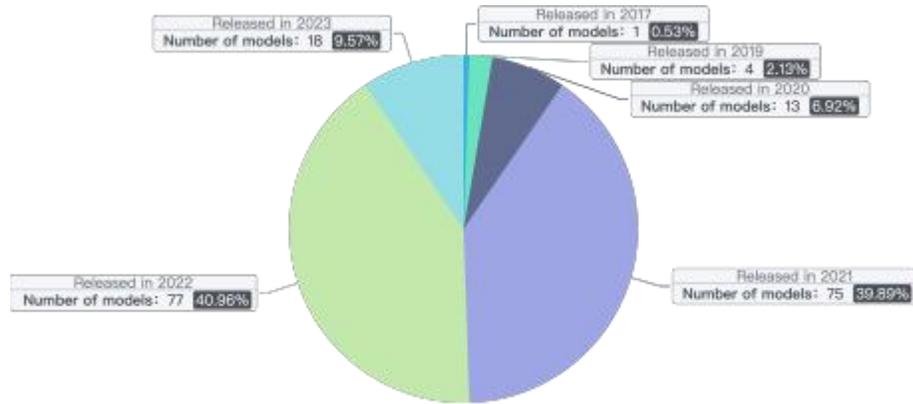

Fig. 1: Proportional representation of model releases with parameters exceeding 1 billion, sourced from OpenBMB [34].

The potential applications of LLMs are vast and continue to grow as the technology matures. From Fig. 1, it's evident that the number of models with parameters exceeding 1 billion has seen a significant increase since 2019. This trend can be traced back to a sequence of developments and key events. The stage was set in 2019, highlighted by Microsoft's one billion investment in OpenAI, which underscored the substantial commercial interest in AI and boosted resources for AI development [11]. The momentum continued into 2020 with the launch of NVIDIA's A100 Tensor Core GPU [26], providing researchers with unprecedented computational power and enabling more complex model training. Soon after, the release of OpenAI's GPT-3 model served as a powerful demonstration of the potential of large models [44], spurring further research and development in the field. As we transitioned into 2021, innovative training techniques like mixed precision training, model parallelism, and data parallelism became mainstream, facilitating the training of even larger models. Concurrently, despite these technological advancements and the evident potential of large-scale models, their rapid proliferation has also raised concerns within the academic community, particularly with regard to the security implications of these models [52]. These concerns underscore the necessity of ongoing research into secure and ethical practices in the development and deployment of LLMs.

Researchers have conducted literature reviews focusing on the security risks and evaluation methods for LLMs. Weidinger et al. (2022) in Google DeepMind outlined a comprehensive taxonomy of ethical and social risks associated with large-scale language models, providing a useful framework to guide responsible innovation in this space[56]. However, it lacks a specific implementation plan necessary for conducting a security evaluation. Chang et al. (2023) presented a review of research work on the evaluation of LLMs[6]. They detailed a classification and summary based on three dimensions: "what to evaluate," "where to evaluate," and "how to evaluate." They summarized various successful and unsuccessful cases of LLMs across multiple domains, 28 popular benchmarks for LLM evaluation, and two primary evaluation methods: automatic and manual. But their evaluation focused more on benchmarks for evaluating LLMs, without comprehensive analysis of the evaluation steps. Huanget al. (2023) have provided a review of the various known vulnerabilities of LLMs, considering both technical methods and ethical norms to promote the safety and credibility of LLMs[18]. Their work serves as a reference for the further development of LLMs. Still, their discussion of known evaluation frameworks is not comprehensive. Glukhov et al. (2023) overviewed various aspects of enhancing the security and trustworthiness of LLMs, including validation, monitoring, and ethical norms, and discussed these from a verification and validation perspective[15], but their research mainly focused on the evaluation of prompt injection risks. Zhao et al. (2023) conducted a systematic and comprehensive survey of LLMs, covering various aspects such as pre-training, fine-tuning, utilization, and evaluation[63]. They also summarized related resources, experimental analyses, and prompt design guidelines for LLMs. Nevertheless, they have not categorized security evaluation frameworks, and have not conducted an analysis to classify the evaluation processes into categories. Existing review papers have covered certain aspects of security risks and evaluation methods for LLMs. On the other hand, a review that comprehensively covers the latest advancements in security evaluation techniques for LLMs is still lacking. Thus this paper will provide such a comprehensive review.

This literature review focuses on scholarly articles related to the security evaluation for LLMs, primarily sourced from two academic search engines: arXiv and Google Scholar. Our search parameters were confined to works published from the year 2020 to the present. The selection criteria for the articles included in this review were based on their relevance to the safety evaluation of LLMs. The selected articles were then thoroughly analyzed, synthesized, and reported in this review. The key contributions of this paper areas follows:

1. We conduct an in-depth study of the security evaluation frameworks for LLMs, distinguishing them into two categories: black-box frameworks and white-box frameworks. Additionally, we provide typical application examples of these frameworks.
2. We abstract Data-driven Evaluation, Attack Simulation, and Formal Verification as three types of security evaluation processes for LLMs, and we analyze application cases for each of these distinct processes.
3. We discuss the underexplored area of security evaluation for multimodal LLMs, the increasing demand for automatic security evaluation systems, and the broad range of vulnerabilities LLMs can fall prey to. Furthermore, we propose future work to focus on the development of automated evaluation platforms, comprehensive coverage of threats, and dedicated research into the unique vulnerabilities and security complexities of multimodal LLMs.

Here is a brief overview of the article's organization. This paper begins in Section 2 by introducing commonly used evaluation metrics in the safety assessment of large language models, including general metrics, metrics for dialogue-oriented language models, metrics for code-generation language models, and metrics for multimodal language models. Following this, Section 3 classifies the safety evaluation framework of large language models into two main categories: black-box evaluation frameworks and white-box evaluation frameworks, and elaborates on the hierarchical structure of each type of framework. Section 4 then categorizes the safety evaluation process of language models from a granular perspective into three types: data-driven evaluation, attack simulation, and formal verification, while providing examples of typical application scenarios for each type of evaluation process. Finally, Section 5 summarizes the main viewpoints of this study, namely, that there is a relative lack of research on the safety evaluation of multimodal language models, a scarcity of fully automated safety evaluation platforms, and inadequate coverage of threats to language models, etc. It also envisages future research directions from the development of automated evaluation platforms, comprehensive coverage of threats, and dedicated research on multimodal language models.

## 2   Evaluation Metrics

In this section, we introduce some common metrics used in the security evaluation of large models and discuss how they are utilized. We have delved into the evaluation indicators that measure the safety of LLMs. These metrics are crucial in determining the effectiveness of a model, its potential risks, and its ability to handle adversarial attacks. Theoretical analysis of these indicators has provided a basis for understanding the different aspects that influence a model's safety.

### 2.1   General Metrics

General Metrics refer to evaluation metrics that provide a general indication of the models' capabilities without being tailored to any specific type of LLMs. Here are some key examples.

- **F1 Score:** F1 score is a harmonic mean of precision and recall, combining them into a single metric that balances both [49]. It evaluates the predictive performance of classification models, ranging from 0 to 1 with 1 being the best. F1 score is useful for imbalanced datasets as it examines precision and recall together. To illustrate, the F1 score serves as a tool to evaluate the performance of classifiers for large language models in detecting unsafe text. A higher F1 score signifies, for instance, that the classifier can detect unsafe text more accurately.
- **Perplexity:** Perplexity (PPL) is a commonly used model evaluation metric in fields such as natural language processing and speech recognition. Perplexity is typically used to assess the quality of a language model, that is, the model's ability to predict a piece of text [39,58]. The definition of Perplexity is the average branching factor when a given probabilistic model predicts a sample set. In the context of language models, the branching factor can be understood as the model's uncertainty in predicting the next word at a certain point in time. The smaller the Perplexity value, the better the performance of the model. By way of example, by testing the Perplexity of large language models for malicious inputs, it becomes clear that a higher Perplexity means the model is more uncertain when dealing with malicious inputs, thus enhancing its reliability from a safety perspective.
– **Attack Success Rate:** Attack Success Rate (ASR), calculated using an equation involving the attacked model, test sample, its target label, and an indicator function, indicates the fraction of successful to total attacks [29,13]. Higher ASR denotes increased security risk due to potential backdoor exploitation. As an example, backdoor attacks can be executed on large language models and the proportion of successful attacks can be calculated to assess the security of the model. A lower ASR demonstrates, for instance, that the model has a stronger ability to withstand backdoor attacks.

### 2.2   Metrics for Dialogue-Oriented LLMs

Metrics for Dialogue-Oriented LLMs refer to evaluation metrics tailored for assessing the conversational abilities of LLMs designed for dialogue applications. These metrics aim to evaluate the natural, coherent during conversations. Some key examples areas following.

– **Coherence:** Measures how coherent and logically consistent a conversational response is [8,25,47]. Evaluation can be either automatically or through human rating. For example, evaluators have the opportunity to score the malicious dialogue responses generated by the language model. A lower Coherence score implies, for example, that the model's response to malicious inputs is less coherent, and hence safer.
– **Grammatical Errors:** Grammatical Errors are the number of errors in the adversarial example's grammar using LanguageTool [47]. LanguageTool is an open-source natural language processing tool that can detect grammatical errors in multiple languages[50], including English, German, French, etc. It can identify and mark typos, punctuation errors, breaches of grammatical rules, and inappropriate usage of language in the text. Counting the number of grammatical errors in adversarial examples can help us understand the robustness of a model, that is, the model's ability to handle erroneous inputs. For a high-quality language processing model, it should be able to provide relatively accurate and useful outputs even when faced with inputs that contain a large number of grammatical errors. Conducting such tests on the model can help us identify areas of the model that need to be improved or fixed. In illustration, intentionally grammatically incorrect malicious inputs can be fed into the language model, with the number of grammatical errors in the output counted. The presence of more errors indicates, for instance, that the model has not effectively processed the malicious input, thereby increasing safety.

### 2.3   Metrics for Code Generation-Oriented LLMs

Metrics for Code Generation-Oriented LLMs refer to the metrics of measuring and evaluating LLMs that are intended to be used for automatic code generation [30,43].

– **Functional Correctness:** Functional Correctness refers to whether a program fulfills its intended functionality and behaves as expected for all possible inputs [31]. For example, HUMANEVAL contains programming problems with ground truth solutions and a limited set of test cases [19,54]. To evaluate an LLM's code generation capability, its synthesized code snippets are run against the test cases in HUMANEVAL to check if the outputs match those of the ground truth solutions. To put it in perspective, it's possible to determine whether the code generated by the language model can pass functional correctness tests in security-related scenarios, thus forming an evaluation of its safety.

### 2.4   Metrics for Multimodal LLMs

Metrics for Multimodal LLMs refer to evaluation metrics designed specifically for assessing the performance of LLMs that process and generate multimodal outputs spanning text, images, audio, and video.

– **Perceptual ability:** Perceptual ability evaluation metrics measure how well Multimodal LLMs can understand and reason about perceptual concepts like images, audio, and video [6]. These metrics test the model's ability to generate accurate descriptions, answer questions, and make inferences about perceptual inputs. Higher scores indicate the model has better perceptual abilities and understanding. Lower scores suggest the model struggles with perceiving and reasoning about multimodal inputs. For instance, testing the multimodal language model's ability to comprehend visual inputs with malicious intent can be informative. A lower Perceptual Ability suggests, for instance, that the model is less likely to understand malicious visual inputs, thereby making it safer.

## 3   Evaluation Frameworks

In the field of security evaluation for LLMs, it is crucial to have a system for evaluating the performance and safety of these models. An evaluation framework for LLMs refers to a comprehensive set of testing methodologies, metrics, and procedures designed to thoroughly assess the capabilities and behaviors of LLMs. Based on the literature reviewed, we explore the security evaluation frameworks that have been developed and implemented, categorize these frameworks into two types according to the traditional black-box/white-box evaluation, and abstract each type of framework into three layers: a user interaction layer, an LLM layer, and an evaluation benchmark layer. The security evaluation frameworks, as depicted in table. 1, can be employed to assess a range of aspects, and provide standardized benchmarks that allow standardized comparisons across different LLMs. By thoroughly evaluating LLMs across these critical dimensions, an effective framework provides comprehensive insight into model strengths, weaknesses and overall readiness for real-world deployment.

Table 1: Typical Applications for Black Box/White Box Security Evaluation Frameworks for LLMs.

| Applications | Categories | Description |
| --- | --- | --- |
| Security and robustness assessments [38,39,60,42]<br><br>Functional correctness evaluation [37,7,43,31,21]<br><br>Downstream performance benchmarking [33,47]<br><br>Human studies [45,10]<br><br>Safety and ethics evaluation [28,59,51] | Black-box Evaluation Framework | For example, [51] have developed a security evaluation system for Chinese language models. Their work investigates the comprehensive security of these models across two dimensions: eight typical security scenarios and six challenging command attack types. The typical security scenarios encapsulate eight prevalent types of security concerns: insults, unfairness and discrimination, crime and illegal activities, sensitive topics, physical harm, mental health, privacy and property, and ethics and morality. The command attacks involve six types of directive input attacks: target hijacking, prompt leakage, role-playing commands, unsafe command topics, queries with unsafe views, and reverse exposure. Using this security evaluation system, they have conducted security evaluations on 15 well-known Chinese language models, including those from the GPT series. |
| Representation analysis [29,62,40,53,46]<br><br>Behavior characterization [19,19,24,20]<br><br>Model inversion [61,17,48] | White-box Evaluation Framework | For example, [48] proposed a comprehensive framework that employs model evaluation as a tool for identifying and mitigating extreme risks. This framework critically examines the potential hazards, likelihood of misalignment, and the agency of LLMs from various perspectives. Importantly, this rigorous examination is conducted at all stages of the model lifecycle, including pre-training, post-training, pre-deployment, and post-deployment. |

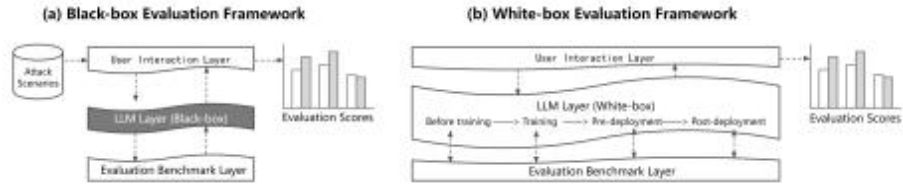

Fig. 2: LLMs evaluation frameworks, which include: (a) black-box evaluation framework, (b) white-box evaluation framework.

### 3.1 Black-box Evaluation Framework

In such a security evaluation framework, as shown in Fig. 2 (a), the LLM itself is regarded as a blackbox, obtaining feedback through interaction with external tools, achieving self-verification and self-improvement [9,22].

**User Interaction Layer:** This layer focuses on interacting with users and collecting test prompts. It uses public datasets as input or designs prompts covering different safety scenarios and instruction attacks to test the safety issues of LLMs. The prompts are designed to be more likely to trigger unsafe responses.

**LLM Layer:** This layer utilizes the pre-trained LLMs to generate responses given the prompts collected from the user interaction layer. Different LLMs can be evaluated and their safety issues can be exposed based on their generated responses.

**Evaluation Benchmark Layer:** This layer evaluates the safety of the generated responses using an LLM evaluator. It compares the prompt and response to judge whether the response is safe or not. The safety scores are then calculated and ranked on a public leaderboard, or a report is generated for response. This layer provides an automatic/semi-automatic and standardized way to assess the safety of LLMs.

### 3.2 White-box Evaluation Framework

White-box Evaluation Framework regards the language model itself as transparent, rather than a black box. Instead of solely relying on external tools for verification and improvement, this framework enables internal processes of self-evaluation [33]. By introspecting on its own knowledge, reasoning, and behavior, the model can identify areas needing improvement. In the White-box Evaluation Framework, we have additional visibility into the internal mechanisms of the LLM compared to the black-box framework as shown in Fig. 2 (b). This enables more rigorous testing and verification of model safety. The three layers areas follows.

**User Interaction Layer:** Similar to the black-box framework, this layer focuses on collecting user commands and feedback on test results. However, in white-box testing, user interaction is not limited to entering pre-designed evaluation scenarios. Users can also visually modify model parameters and training data. Furthermore, evaluation results before, during, and after training and deployment should be provided in real-time, empowering users to make more informed decisions about improving model safety. Rather than passively receiving prompts, users actively shape the model's internal representations. They fine-tune mechanisms found to be unsafe, and retrain on modified datasets that eliminate harmful biases. This continues iteratively until the desired safety criteria are met. Such transparent, user-driven refinement of the LLM's internals is only possible in the white-box setting.

**LLM Layer:** In addition to generating responses to prompts, the white-box framework grants access to inspect the model's internal states, such as activations and gradients. This facilitates techniques like unit testing, mechanism analysis, and verification of intended behaviors. For example, researchers can trace whether certain goal representations emerge in the model. This transparency also aids in understanding the model's limitations and potential biases, contributing to safer and more trustworthy AI systems.

**Evaluation Benchmark Layer:** As in the black-box case, this layer judges the safety of the LLM's outputs. However, conducting white-box security evaluations on LLMs at different stages is integral to their development and deployment. This process begins even before training, where white-box security assessments on pre-training models help to discover any existing dangerous capabilities or misalignments. These findings provide invaluable guidance for subsequent training. At this stage, targeted training plans can be designed to specifically avoid the emergence of such dangerous capabilities or to improve alignments. The evaluation process continues during the training phase. Regular assessments are conducted on the model to pinpoint any new dangerous capabilities or misalignments that may appear during the training process. The results of these evaluations then inform necessary adjustments to the training methods. If necessary, any problematic training can be terminated or rolled back to ensure the model's security. Prior to deployment, a comprehensive white-box security evaluation of the model is conducted to determine its readiness for deployment. This assessment helps to identify key risk areas that need to be addressed before the model can be safely deployed. The evaluation results can also inform the establishment of deployment restrictions, such as defining the scope of use or access levels. Once the model is deployed, the evaluation process doesn't stop. It is essential to monitor for any new issues that may arise in the real-world application of the model. User feedback is collected to continually improve the evaluation plan. This feedback loop allows for adjustments to the model or the setting of usage restrictions based on newly discovered issues. Furthermore, post-deployment monitoring is crucial to verify the effectiveness of the security measures that have been put in place.

## 4  Evaluation Processes

Security evaluation process for LLMs refers to the specific application process of the evaluation system built based on the evaluation framework. We categorize the processes of evaluating the security of LLMs according to the granularity of evaluation, ranging from coarse-grained to fine-grained. The table 2 offers a comprehensive overview of three processes used to evaluate LLMs, illustrated with selected typical cases for each process.

Table 2: Typical use cases for security evaluation processes for LLMs.

| Category | Description | Reference |
|---|---|---|
| Data-driven Evaluation | Build the RealToxicityPrompts dataset for evaluating toxicity generation in language models[14]. | Gehman et al. (2020) |
| | Build a dataset containing approximately 40,000 red-blue adversarial dialogues, which can be used to analyze different types of potential risks, establish automated adversarial testing, and more[13]. | Ganguli et al. (2022) |
| | Build the SafeText dataset for evaluating language models' awareness of physical harm[28]. | Levy et al. (2022) |
| | Build the Latent Jailbreak dataset for evaluating text safety and output robustness of language models[42]. | Qiu et al. (2023) |
| | Build the LLMSecEval dataset for security evaluations of language models[54]. | Tony et al. (2023) |
| Attack Simulation | Choose GitHub Copilot as the evaluation target, use MITRE's CWE vulnerability database, design 89 different scenarios, and evaluate the vulnerability risks of GitHub Copilot generating code in different security-related scenarios[37]. | Pearce et al. (2022) |
| | Construct a prompt dataset for CHATGPT jailbreak, and comprehensively evaluate the ability and robustness of different jailbreak prompts in bypassing CHATGPT restrictions[32]. | Liu et al. (2023) |
| | Construct image, audio, and dialogue datasets for the two open-source multimodal language models, LLaVA and PandaGPT[3]. | Bagdasaryan et al. (2023) |
| | Test the ability of ChatGPT to generate malicious code, phishing emails, 0day attacks, macros, and Living Off The Land Binaries (LOLBINs)[41]. | Qammar et al. (2023) |
| Formal Verification | Define privacy rules such as not disclosing identity information and not memorizing training data, and use automated methods to check model outputs against these predefined rules[36]. | Pan et al. (2020) |
| | Quantitatively define the concept of "memory" in LLMs, that is, text fragments that only appear in a small number of training samples, and design an experiment to extract private information[5]. | Carlini et al. (2021) |
| | Formalize the task of auditing LLMs as a discrete optimization problem, and optimize from random initialization of inputs and outputs on language models like GPT-2, until finding input-output pairs that satisfy the objective function and constraints[22]. | Jones et al. (2023) |

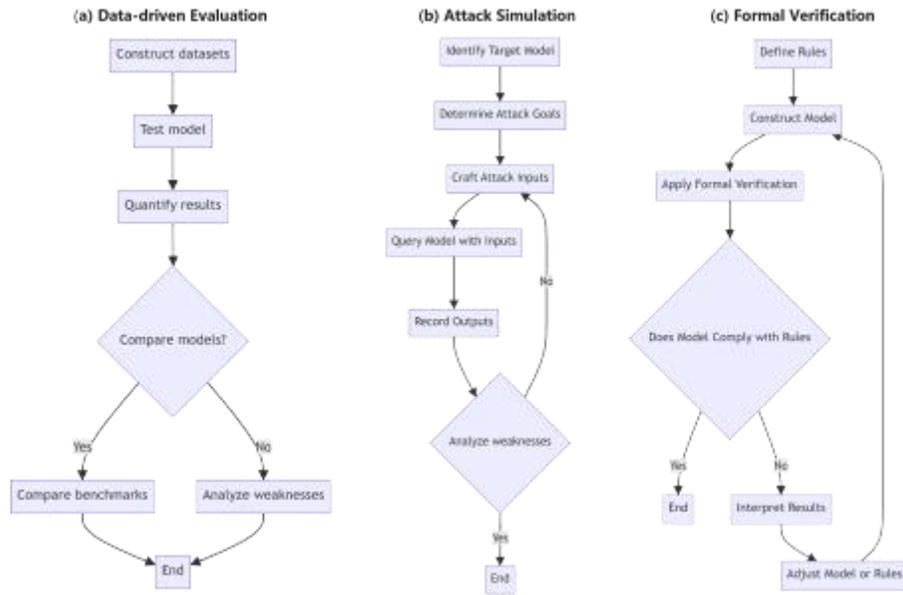

Fig. 3: Three major language model security evaluation processes, which include: (a) data-driven evaluation, (b) attack simulation, (c) formal verification.

The different processes include data-driven evaluation, attack simulation, and formal verification, each with its unique applications and nuances. Data-driven evaluation is a versatile approach not restricted to the evaluation of a specific model or a set of models. This process is characterized by the universality of its evaluation datasets, making it suitable for broad applications across different types of models. Attack simulation, in contrast, is a more targeted process where the evaluation dataset is chosen based on the specific model and evaluation objectives. This approach allows for a more personalized evaluation, specifically tailored to the nuances of the model under scrutiny. Formal verification is primarily focused on the evaluation of fine-grained tasks and iterative optimization of models. This process facilitates a deeper understanding of the model's performance on specific tasks, and aids in guiding the model's development and improvement.

Despite the nuances of each process, they are not mutually exclusive. Instead, they can be transformed and adapted according to different evaluation scenarios. This allows for a comprehensive, multi-faceted evaluation of LLMs, ensuring a more accurate understanding of their functioning and performance, and improving LLMs' security.

### 4.1 Data-driven Evaluation

Data-driven Evaluation refers to a prevalent approach which is creating tailored datasets to probe known vulnerabilities [42,53]. For example, REALTOXICITYPROMPTS consists of 100K naturally occurring toxic prompts to test if LLMs generate harmful text [14]. Adversarial NLI builds sentence pairs to evaluate an LLM's ability to detect contradictions. By crafting test cases that require reasoning about security, data-driven evaluation offers a scalable way to benchmark model security. However, dataset biases may skew conclusions [27]. The main steps of Data-driven Evaluation as shown in Fig. 3 (a) are:

(1) Construct customized datasets to expose vulnerabilities.
(2) Test LLMs on datasets and observe outputs.
(3) Quantify security metrics like attack accuracy rate.
(4) Compare with other LLMs.
(5) Analyze and identify model weaknesses.

### 4.2 Attack Simulation

Attack Simulation refers to testing LLMs through simulated attacks to evaluate their security and robustness. Researchers design attacks meant to trick, mislead, or exploit weaknesses in LLMs in order to assess vulnerabilities. Researchers directly perform attacks by querying LLMs with carefully designed inputs [16,23,38,64]. For instance, targeted misspellings and paraphrasing help circumvent blocking of dangerous keywords. Multi-turn conversations reveal risks of extracted information accumulation. Attack simulation provides direct evidence of vulnerabilities. But new defenses may impede replication. Documenting details of the experimental setup is crucial.

Inspection-based Analysis inspection of model parameters, training data, and generated texts can reveal security issues. Analyzing correlations between training data artifacts and model outputs can expose unintended memorization. Manual inspection offers intuitions into failure modes. But it relies on expert judgments and small samples. Developing rigorous auditing protocols is an open challenge. Here are key steps in Fig. 3 (b):

(1) Identify target LLM to evaluate
(2) Determine attack goals and scenarios, e.g. extracting sensitive messages from training data.
(3) Craft attacks using carefully designed LLM inputs.
(4) Query the LLM with attack inputs and record outputs.
(5) Analyze if attack goals were achieved based on outputs.
(6) Iterate attack approaches to improve effectiveness.
(7) Draw conclusions on LLM vulnerabilities.

### 4.3  Formal Verification

Formal Verification serves as a methodology for inspecting whether a system adheres to a set of predefined specifications [18,48]. In the context of evaluating the safety of LLMs, Formal Verification can be employed to ensure the behavior of the model aligns with predetermined rules or standards [12,20]. Consider an instance where we desire to validate the LLM's behavior when it processes sensitive information. Rules can be established, such as the model should not generate outputs that could potentially disclose personal identifying information when given such inputs. Subsequently, we apply Formal Verification to examine whether the model complies with these rules. The key steps as shown in Fig. 3 (c) are:

(1) Identify the rules or standards you wish the model to observe. These should be designated based on your specific application and the safety concerns you are addressing.

(2) Constructing model, which typically involves training the LLM.

(3) Implement Formal Verification techniques to inspect whether your model complies with the rules defined in the first step.

(4) Based on the results of Formal Verification adjust the model or the rules defined in the first step.

(5) Repeat the above steps as needed, until the LLM satisfies all rules.

## 5  Conclusions and Perspectives

This paper has provided a comprehensive analysis of evaluation metrics, evaluation frameworks, and evaluation processes for LLMs. Here are the main findings.

- Comparatively, research into multimodal LLMs is less prevalent than in text and code models, not only in terms of defining evaluation metrics but also in the design and development of evaluation frameworks. This discrepancy might be due to the smaller size and utilization of multimodal models relative to other types.
- Fully automated platforms for security evaluation of LLMs are rarely released. However, given the increasing number of applications based on LLMs, the demand for automatic security evaluation is growing day by day.
- Pertaining to the types of vulnerabilities in the evaluation, LLMs are prone to a wide range of threat vulnerabilities. Even when the scope is limited to OWASP Top 10 for LLMs, which is a list of the top 10 security risks for LLM applications proposed by the Open Web Application Security Project (OWASP) [35], no current security evaluation framework can completely cover and verify all of them.

Given the current state of research, there is a clear need for more dedicated attention to security threats and evaluations of LLMs. Future work should focus on the following possible directions.

- **Development of Automated Evaluation Platforms:** There is a pressing need for the development and release of fully automated platforms for LLMs' security evaluation. These platforms could help streamline the process of identifying vulnerabilities in LLMs and thus, contribute to their safe and secure deployment.
- **Comprehensive Coverage of Threats:** Future research should aim at developing evaluation frameworks that can cover a wider range of threat vulnerabilities. It is crucial to consider not only the OWASP Top 10 for LLMs but also other potential threats that might emerge with the evolution of LLMs and their applications.
- **Emphasize the area of multimodal LLMs:** Given the limited research on the security evaluation of multimodal LLMs to date, future research should emphasize this area. Exploring the development of mature security evaluation systems for multimodal LLMs and their specific vulnerabilities could provide a more holistic understanding of the security complexities involved with LLMs.

In conclusion, while significant strides have been made in understanding the security aspects of LLMs, there is still much to learn. It is hoped that this work will stimulate further research in this critical domain.

## Acknowledgements

We would like to acknowledge the role of AI and AI-assisted writing technologies, specifically the Claude2 and GPT-4 models, in refining the grammar of this manuscript.

In addition to technological support, this research was funded by the Major Key Project of PCL (Grant No. PCL2022A03-3), NSFC (No. 62072131, 61972106), Key R&D Program of Guangdong Province (No.2019B010136003), Guangdong Basic and Applied Basic Research Foundation (No.2022A1515011401), DongGuan Innovative Research Team Program (No. 2018607201008).

16      J. Huang et al.

Exploring Advanced Methodologies in Security Evaluation for LLMs    1734. OpenBMB: BMList. https://openbmb.github.io/BMList/ (Aug 2023), (Accessed 7 September 2023)
35. OWASP: OWASP Top 10 for Large Language Model Applications. https://owasp.org/www-project-top-10-for-large-language-model-applications/ (Aug 2023), (Accessed 7 September 2023)
36. Pan, X., Zhang, M., Ji, S., Yang, M.: Privacy Risks of General-Purpose Language Models. In: 2020 IEEE Symposium on Security and Privacy (SP). pp. 1314–1331 (May 2020). https://doi.org/10.1109/SP40000.2020.00095
37. Pearce, H., Ahmad, B., Tan, B., Dolan-Gavitt, B., Karri, R.: Asleep at the Keyboard? Assessing the Security of GitHub Copilot's Code Contributions. In: 2022 IEEE Symposium on Security and Privacy (SP). pp. 754–768 (May 2022). https://doi.org/10.1109/SP46214.2022.9833571
38. Pedro, R., Castro, D., Carreira, P., Santos, N.: From Prompt Injections to SQL Injection Attacks: How Protected is Your LLM-Integrated Web Application? (2023). https://doi.org/10.48550/ARXIV.2308.01990
39. Perez, F., Ribeiro, I.: Ignore Previous Prompt: Attack Techniques For Language Models (2022). https://doi.org/10.48550/ARXIV.2211.09527
40. Pezeshkpour, P.: Measuring and Modifying Factual Knowledge in Large Language Models (2023). https://doi.org/10.48550/ARXIV.2306.06264
41. Qammar, A., Wang, H., Ding, J., Naouri, A., Daneshmand, M., Ning, H.: Chatbots to ChatGPT in a Cybersecurity Space: Evolution, Vulnerabilities, Attacks, Challenges, and Future Recommendations (2023)
42. Qiu, H., Zhang, S., Li, A., He, H., Lan, Z.: Latent Jailbreak: A Benchmark for Evaluating Text Safety and Output Robustness of Large Language Models (2023). https://doi.org/10.48550/ARXIV.2307.08487
43. Ren, S., Guo, D., Lu, S., Zhou, L., Liu, S., Tang, D., Sundaresan, N., Zhou, M., Blanco, A., Ma, S.: CodeBLEU: A Method for Automatic Evaluation of Code Synthesis (2020). https://doi.org/10.48550/ARXIV.2009.10297
44. Sagar, R.: OpenAI Releases GPT-3, The Largest Model So Far. https://analyticsindiamag.com/open-ai-gpt-3-language-model/ (Jun 2020), (Accessed 7 September 2023)
45. Sandoval, G., Pearce, H., Nys, T., Karri, R., Garg, S., Dolan-Gavitt, B.: Lost at C: A User Study on the Security Implications of Large Language Model Code Assistants (2022). https://doi.org/10.48550/ARXIV.2208.09727
46. Shao, H., Huang, J., Zheng, S., Chang, K.C.C.: Quantifying Association Capabilities of Large Language Models and Its Implications on Privacy Leakage (2023). https://doi.org/10.48550/ARXIV.2305.12707
47. Shen, X., Chen, Z., Backes, M., Zhang, Y.: In ChatGPT We Trust? Measuring and Characterizing the Reliability of ChatGPT (2023). https://doi.org/10.48550/ARXIV.2304.08979
48. Shevlane, T., Farquhar, S., Garfinkel, B., Phuong, M., Whittlestone, J., Leung, J., Kokotajlo, D., Marchal, N., Anderljung, M., Kolt, N., Ho, L., Siddarth, D., Avin, S., Hawkins, W., Kim, B., Gabriel, I., Bolina, V., Clark, J., Bengio, Y., Christiano, P., Dafoe, A.: Model evaluation for extreme risks (2023). https://doi.org/10.48550/ARXIV.2305.15324
49. Stiff, H., Johansson, F.: Detecting computer-generated disinformation. International Journal of Data Science and Analytics **13**(4), 363–383 (May 2022). https://doi.org/10.1007/s41060-021-00299-5
50. Stratus-Security: FinGen - Penetration Testing Findings Generator. https://github.com/Stratus-Security/FinGen (Aug 2023), (Accessed 7 September 2023)